\documentclass[unsortedaddress, aps, prd, reprint, nofootinbib]{revtex4-2}

\usepackage{graphicx}% Include figure files
\usepackage{amsmath,amssymb,mathtools}
\usepackage{dcolumn}% Align table columns on decimal point
\usepackage{bm}% bold math
\usepackage{xcolor}
\usepackage{soul}
\usepackage{braket}
\usepackage[separate-uncertainty=true,multi-part-units=single]{siunitx}
\usepackage{scalerel}
\usepackage{enumitem}

\definecolor{Mathematica1}{rgb}{0.368417, 0.506779, 0.709798}
\definecolor{Mathematica2}{rgb}{0.880722, 0.611041, 0.142051}
\definecolor{Mathematica3}{rgb}{0.560181, 0.691569, 0.194885}
\definecolor{darkred}{rgb}{0.545,0,0}
\definecolor{dullblue}{rgb}{0,0.298,0.49}
\definecolor{blue3}{RGB}{31, 119, 180}
\usepackage[colorlinks=true
,urlcolor=dullblue
,anchorcolor=blue3
,citecolor=dullblue
,filecolor=blue3
,linkcolor=darkred
,menucolor=blue3
,pagecolor=blue3
,linktocpage=true
,pdfproducer=medialab
]{hyperref}

\newcommand{\dd}{\mathop{\mathrm{d}\!}{}}

\newcommand{\slab}[1]{{\textsc{#1}}}

\newcommand{\minus}{{\protect \scalebox {0.75}[1.0]{$-$}}}

\renewcommand{\vec}[1]{\boldsymbol{\mathbf{#1}}}

\DeclareMathOperator\diag{diag}

\newcommand{\ped}[1]{_{\mathrm{#1}}}
\renewcommand{\ap}[1]{^{\mathrm{#1}}}
\DeclarePairedDelimiter{\abs}{\lvert}{\rvert}

\def\beq{\begin{equation}}
\def\eeq{\end{equation}}

\begin{document}

\title{Smooth binary evolution from wide resonances in boson clouds}
% Force line breaks with \\

\author{Giovanni Maria Tomaselli}

\affiliation{School of Natural Sciences, Institute for Advanced Study, Princeton, NJ 08540, USA}

\begin{abstract}

Ultralight scalars can form superradiant clouds around rotating black holes. These may alter the dynamics of compact binaries and the ensuing waveform through orbital resonances and cloud ionization. We re-examine resonances involving states with nonzero decay width, deriving an effective treatment for resonances that are wider than the binary's frequency chirp. We demonstrate the utility of this approach by calculating an upper bound for the cloud's mass surviving up to the latest stages of the inspiral. Next, we study the accumulation of resonances with high-energy bound states. When these infinitely many, increasingly weak resonances are properly taken into account, they smooth out the ``sharp features'' in the binary's evolution that had been attributed to the ionization of the cloud. We compare our Newtonian results with recent relativistic calculations, highlighting common features as well as discrepancies. Our conclusions emphasize the need to carefully incorporate resonances in boson cloud waveform modeling.

\end{abstract}

\maketitle

\section{Introduction}

After almost ten years from the first detection of gravitational waves (GWs) \cite{LIGOScientific:2016aoc}, hundreds of compact object mergers have been detected by the LIGO-Virgo-KAGRA collaboration \cite{LIGOScientific:2018mvr,LIGOScientific:2020ibl,KAGRA:2021vkt}. While the discovery of stellar-mass objects proceeds steadily with current interferometers, several low-frequency detectors are expected to come online in about a decade from now. These include ground-based detectors, such as the Einstein Telescope \cite{Abac:2025saz} and Cosmic Explorer \cite{Evans:2021gyd}, and space-borne ones, such as LISA \cite{LISA:2024hlh}, TianQin \cite{TianQin:2020hid}, and Taiji \cite{Hu:2017mde}. One of the main scientific targets of these missions is the slow inspirals of stellar-mass objects into intermediate-mass or supermassive black holes (BHs), also known as extreme mass ratio inspirals (EMRIs). These systems are particularly well suited to probe the environments around BHs, such as accretion disks \cite{Barausse:2007dy,Kocsis:2011dr,Derdzinski:2020wlw,Speri:2022upm,Cole:2022yzw,Garg:2024oeu}, as the extremely long EMRI waveforms allow environmental effects to build up over time. 

A small mass ratio also means that the binary's perturbation to the environment is small, which makes EMRIs ideal to probe \emph{dark} environments \cite{Eda:2013gg,Eda:2014kra,Macedo:2013qea,Kavanagh:2020cfn,Becker:2022wlo}, which would otherwise get easily disrupted---although certain setups allow them to survive even for equal massses \cite{Bamber:2022pbs,Aurrekoetxea:2023jwk,Aurrekoetxea:2024cqd,Tomaselli:2024ojz}. In this paper, we consider BHs surrounded by superradiant clouds of ultralight scalars \cite{Starobinsky:1973aij,Wilczek:1977pj,Arvanitaki:2009fg,Arvanitaki:2010sy,Brito:2015oca}, also known as ``gravitational atoms.'' These clouds form naturally around rapidly rotating black holes in the presence of a massive boson whose Compton wavelength is of the order of the BH's size.

When a gravitational atom is part of a binary system, a very rich phenomenology arises \cite{Baumann:2018vus,Zhang:2018kib,Berti:2019wnn,Baumann:2019ztm,Takahashi:2021eso}. The two most striking kinds of interaction are orbital resonances \cite{Tomaselli:2024bdd,Tomaselli:2024dbw,Boskovic:2024fga} and dynamical friction (also known as ionization) \cite{Takahashi:2021yhy,Baumann:2021fkf,Baumann:2022pkl,Tomaselli:2023ysb}, though other effects such as accretion \cite{Baumann:2021fkf} and the cloud's self-gravity \cite{Ferreira:2017pth,Hannuksela:2018izj} also proved to be relevant. These phenomena have been discovered using a nonrelativistic approach, where both the cloud and the binary's perturbation are modeled within Newtonian gravity. More recent works have taken one step forward, adopting a fully relativistic formalism \cite{Brito:2023pyl,Dyson:2025dlj,Li:2025ffh}. The main result of these papers is the determination of the scalar flux moving towards infinity and into the BH horizon. The former has been shown \cite{Dyson:2025dlj} to match the Newtonian prediction from ionization. The latter has remained harder to interpret: while generally believed to be due to resonances, no clear prediction for the horizon flux is available from Newtonian works. This situation motivates us to improve the nonrelativistic modeling of resonances.

In this work, we consider resonances with states with large decay width, within the usual nonrelativistic approximations. We derive and discuss the ``resonance power'', a natural candidate to be the Newtonian counterpart of the horizon flux calculated in the relativistic papers. Although closely related quantities have been discussed in the literature \cite{Tong:2022bbl,Takahashi:2023flk,Fan:2023jjj,Zhu:2024bqs,DellaMonica:2025zby}, we study here in detail and for the first time the phenomenological implications of this approach.

We start by demonstrating that previous results, obtained by explicitly solving the Schrödinger equation for the cloud together with the orbital backreaction, can be recovered in an easier and more intuitive way. This includes the phenomenon of ``resonance breaking,'' discovered in \cite{Tomaselli:2024bdd}, which critically determines the resonant history of the cloud-binary system. Calculating the resonance power also allows for a simpler approach to orbits with generic eccentricity and inclination. Armed with these tools, we proceed to make two predictions.

First, we calculate an upper bound for the cloud's mass surviving to the final stages of the inspiral. This is particularly relevant, because the last orbits correspond to the loudest, and therefore easier to detect and analyze, GW signal. This bound comes from the requirement that the binary moves past the last resonance, which involves the lowest-energy spherically symmetric state of the cloud. The upper limit on the cloud's mass only depends on the boson mass and on the orbital inclination.

Second, we study resonances that occur with bound states whose energy approaches zero. These infinitely many, weak and closely spaced resonances accumulate at a frequency where the ionization power is known to have a discontinuity. This feature was claimed in \cite{Tomaselli:2024bdd,Tomaselli:2024dbw} (and later confirmed by other works, such as \cite{Brito:2023pyl,Cao:2024wby}) to produce \emph{sharp features} in the binary's orbital evolution, and the ensuing gravitational waveform. We prove that the combined effect of all these resonances creates a discontinuous jump in the resonance power, which cancels exactly the one from ionization. We conclude therefore that the orbital evolution must be \emph{smooth}, and no sharp features exist. As a consequence, studies of the binary-cloud evolution cannot take into account just a handful of strong resonances---a proper treatment of the infinitely many weak and closely spaced ones is mandatory.

Finally, we compare our results with the relativistic ones from \cite{Brito:2023pyl,Dyson:2025dlj,Li:2025ffh}. The comparison is only partially satisfactory: despite the presence of several common features, there are also large and currently unexplained differences. We speculate on the possible causes.

The paper is organized as follows. In Sec.~\ref{sec:wide-resonances}, we study resonances with nonzero decay width, calculate analytically and numerically the resonance power, and discuss the limitations introduced by a nonconstant orbital frequency. In Sec.~\ref{sec:backreaction}, we demonstrate that we can recover the physics of floating resonances, including their ``breaking,'' using the resonant power as an instantaneous energy loss. In Sec.~\ref{sec:wide-resonance-phenomenology}, we extend the results to generic orbits and derive the upper bound on the cloud's mass surviving up to the merger. In Sec.~\ref{sec:no-sharp-features}, we prove that the ionization discontinuity, and the ensuing sharp features in the orbital evolution, are eliminated when the resonances are properly taken into account. In Sec.~\ref{sec:comparison}, we compare our results with those of \cite{Brito:2023pyl,Dyson:2025dlj,Li:2025ffh}. We conclude in Sec.~\ref{sec:conclusions}.

Throughout the paper we use natural units with $G=c=\hbar=1$. The code used in this work is available on GitHub \cite{Pres:github}.

\section{Wide resonances}
\label{sec:wide-resonances}

\subsection{Binary perturbation}
\label{sec:binary-perturbation}

We consider a setup where a larger BH, with mass $M$, is surrounded by a scalar cloud of mass $M\ped{c}$ in a hydrogenic state $\ket{a}=\ket{n_a\ell_am_a}$. The gravitational fine structure constant is given by $\alpha=M\mu$, where $\mu$ is the scalar's mass. We sometimes refer to the BH-cloud system as a ``gravitational atom.'' We refer the reader to \cite{Baumann:2018vus,Baumann:2019eav} for details on the cloud's energy spectrum and wavefunctions.

A smaller compact object, with mass $qM$, orbits the primary BH with frequency $\Omega$. Throughout the paper we will work at leading order in the mass ratio $q$. The binary's periodic gravitational perturbation is
\beq
V_*(t,\vec r)=-\frac{qM}{\abs{\vec r-\vec R_*(t)}}+\frac{qM}{R_*^3}\vec{R_*\cdot r}\,,
\label{eqn:V_*}
\eeq
where $\vec R_*$ is the compact object's position, and the second term gives rise to the fictitious force in the accelerated primary BH's frame. The perturbation connects states of the cloud with matrix elements
\beq
\braket{a|V_*|b}=\sum_{g\in\mathbb Z}\eta^{(g)}e^{ig\Omega t}\,,
\label{eqn:V_fourier}
\eeq
where the Fourier coefficients $\eta^{(g)}$ can be determined by a suitable integral over radial and angular variables. In particular, each $\eta^{(g)}$ is a sum over the multipolar components of $V_*$, labeled by the angular momentum number $\ell_*$: the dipole $\ell_*=1$, the quadrupole $\ell_*=2$, etc. Their full expression can be found, e.g., in \cite{Tomaselli:2023ysb,Tomaselli:2024bdd}.

\subsection{Fixed frequency}
\label{sec:fixed-frequency}

In this Section we study resonances between states with nonzero decay width. We present a new derivation of some results that were first shown in \cite{Tong:2022bbl,Takahashi:2023flk,Takahashi:2024fyq,DellaMonica:2025zby}. As we discuss throughout, our formulas have some, mostly minor, differences compared to the ones in the literature.

To simplify the problem, we start by considering a two-state system, with wavefunction $\psi=c_a\ket{a}+c_b\ket{b}$, whose off-diagonal coupling oscillates with fixed frequency $g\Omega$. The Schrödinger equation then reads
\beq
i\dot\psi=\mathcal H\psi\,,\qquad\mathcal H=\begin{pmatrix}
E_a+i\Gamma_a & \eta e^{ig\Omega t}\\
\eta^*e^{-ig\Omega t} & E_b+i\Gamma_b
\end{pmatrix}\,,
\label{eqn:schrodinger}
\eeq
where $E_{a,b}$ are the states' nonrelativistic energies, and $\Gamma_{a,b}$ are their widths. For simplicity, we neglect here the cloud's dissipation through GWs \cite{Yoshino:2013ofa}, which generally occurs on timescales longer than a typical inspiral.

This equation can be solved analytically by switching to the ``dressed'' frame with the unitary transformation $\psi=e^{ig\Omega t\diag(1,-1)/2}\psi_D$. The dressed-frame Hamiltonian,
\beq
\mathcal H_D=\begin{pmatrix}
E_a+i\Gamma_a+g\Omega/2 & \eta\\
\eta^* & E_b+i\Gamma_b-g\Omega/2
\end{pmatrix}\,,
\label{eqn:H_D}
\eeq
is time-independent, and has eigenvalues
\beq
\omega_\pm=\frac12(E_a+i\Gamma_a+E_b+i\Gamma_b\pm\tilde E)\,,
\eeq
where $\tilde E=\sqrt{(E_a+i\Gamma_a-E_b-i\Gamma_b+g\Omega)^2+4\abs{\eta}^2}$\,. The exact solution of \eqref{eqn:schrodinger} is then
\begin{align}
\label{eqn:solution-shrodinger-ca}
c_a(t)&=A_+e^{-i(\omega_+-g\Omega/2)t}+A_-e^{-i(\omega_--g\Omega/2)t}\,,\\
\label{eqn:solution-shrodinger-cb}
c_b(t)&=B_+e^{-i(\omega_++g\Omega/2)t}+B_-e^{-i(\omega_-+g\Omega/2)t}\,,
\end{align}
where $A_\pm$ and $B_\pm$ are constants.

We now approximate \eqref{eqn:solution-shrodinger-ca} and \eqref{eqn:solution-shrodinger-cb} in the limit where the perturbation is small. To first order in $\abs{\eta}^2$, we have
\begin{align}
\label{eqn:cat}
c_a(t)&\approx A_+e^{-iE_at}e^{(\Gamma_a+\tilde\Gamma)t}{\kern-0.08em}+{\kern-0.08em}A_-e^{-i(E_b-g\Omega)t}e^{(\Gamma_b-\tilde\Gamma)t}\,,\\
\label{eqn:cbt}
c_b(t)&\approx B_+e^{-i(E_a+g\Omega)t}e^{(\Gamma_a+\tilde\Gamma)t}{\kern-0.08em}+{\kern-0.08em}B_-e^{-iE_bt}e^{(\Gamma_b-\tilde\Gamma)t}\,,
\end{align}
where we defined
\beq
\tilde\Gamma\equiv\frac{(\Gamma_b-\Gamma_a)\abs{\eta}^2}{(E_a-E_b+g\Omega)^2+(\Gamma_b-\Gamma_a)^2}\,,
\label{eqn:Gammatilde}
\eeq
and neglected an $\mathcal O(\abs{\eta}^2)$ correction to the real part of $\omega_\pm$. This analytical solution was also derived in \cite{Tong:2022bbl}, where however the term $(\Gamma_b-\Gamma_a)^2$ in the denominator of \eqref{eqn:Gammatilde} was neglected. As we will see, this term plays an important role in our analysis.

By definition, a cloud that saturated the superradiant instability of state $\ket{a}$ has $\Gamma_a=0$. Let us now focus on the case where the other state is decaying, i.e., $\Gamma_b<0$. Equations \eqref{eqn:cat} and \eqref{eqn:cbt} feature two exponentially decaying terms, $e^{\tilde\Gamma t}$ and $e^{(\Gamma_b-\tilde\Gamma)t}$. To leading order in $\abs{\eta}^2$, we have $\abs{\tilde\Gamma}\ll\abs{\Gamma_b}$, so the first exponential decays much more slowly than the second one, which can then be considered as an initial transient. We can thus write
\beq
\abs{c_a}^2\approx\abs{A_+}^2e^{2\tilde\Gamma t}\,,\qquad\abs{c_b}^2\approx \abs{B_+}^2e^{2\tilde\Gamma t}\,,
\label{eqn:cat-cbt-approx}
\eeq
and the cloud's mass $M\ped{c}=\mu(\abs{c_a}^2+\abs{c_b}^2)$ decays according to
\beq
\dot M\ped{c}=2M\ped{c}\tilde\Gamma\,.
\label{eqn:Mdot-one-state}
\eeq
By substituting \eqref{eqn:cat-cbt-approx} into \eqref{eqn:schrodinger}, we can also express the amplitude of state $\ket{b}$ as a function of that of state $\ket{a}$:
\beq
\abs{c_b}^2=\frac{\abs{\eta}^2}{(E_a-E_b+g\Omega)^2+\Gamma_b^2}\abs{c_a}^2\,.
\label{eqn:b-amplitude}
\eeq
Formulae \eqref{eqn:Mdot-one-state} and \eqref{eqn:b-amplitude} were also found in \cite{Takahashi:2023flk} through an ``adiabatic elimination'' of state $\ket{b}$. We notice that these expressions have the typical form of a Lorentzian: close to the resonance condition $E_b-E_a=g\Omega$, the perturbation is particularly efficient at depleting the cloud. Conversely, the effect becomes negligible away from the resonance, whose width is set by $\Gamma_b$.

Compared to the model \eqref{eqn:schrodinger}, multiple complications arise in real systems. The easiest ones to deal with are the presence of more than two states, and a non-monochromatic perturbation. Because the occupancy number $\abs{c_b}^2$ is $\mathcal O(\abs{\eta}^2)$, as shown in \eqref{eqn:b-amplitude}, the coupling between two decaying states will generate corrections of $\mathcal O(\abs{\eta}^4)$ to the population $\abs{c_a}^2$ of the initial state. We can thus safely ignore these corrections and simply sum the contributions of various decaying states to \eqref{eqn:Mdot-one-state}. When the perturbation contains multiple Fourier components, as in \eqref{eqn:V_fourier}, we notice that no two of them can be close to resonance at the same time, as $\abs{\Gamma_b}\ll\abs{E_b-E_a}$ in all practical cases. We can thus treat them independently, approximating the total effect as the sum of their contributions. We thus define the total cloud mass loss rate as
\beq
\dot M\ped{c}\ap{res}=\sum_{\ket{b}}\sum_g\frac{2M\ped{c}\Gamma_b\abs{\eta}^2}{(E_a-E_b+g\Omega)^2+\Gamma_b^2}\,,
\label{eqn:Mdot}
\eeq
where the coupling $\eta$ implicitly depends on $\ket{b}$, $g$ and $\Omega$.

While \eqref{eqn:Mdot} quantifies the impact of the perturbation on the cloud, we are interested in many cases in its orbital backreaction. We thus define the \emph{resonance power} as the rate of change of the cloud's energy, which for a two-state system is
\beq
P\ped{res}=E_a\frac{\dd}{\dd t}\abs{c_a}^2+E_b\frac{\dd}{\dd t}\abs{c_b}^2-2\Gamma_bE_b\abs{c_b}^2\,.
\eeq
Here, the corrective term $-2\Gamma_bE_b\abs{c_b}^2$ ensures that we only include in $P\ped{res}$ the energy change induced by the perturbation, and not that due to the natural decay of $\ket{b}$. Using the Schrödinger equation \eqref{eqn:schrodinger} to expand the time derivatives, we can rewrite $P\ped{res}$ as
\beq
P\ped{res}=\frac{M\ped{c}}\mu\sum_{\ket{b}}\sum_g\frac{2(E_a-E_b)\Gamma_b\abs{\eta}^2}{(E_a-E_b+g\Omega)^2+\Gamma_b^2}\,,
\label{eqn:Pres}
\eeq
where we once again summed over different states and overtones. Note that the quantity $P\ped{res}$ was first introduced in \cite{DellaMonica:2025zby}, where however a factor $-g\Omega$ was used in place of $E_a-E_b$. The two formulas only agree at resonance, when $g\Omega=E_b-E_a$.

\subsection{Numerical results}
\label{sec:numerical-results}

\begin{figure*}
\centering
\includegraphics[]{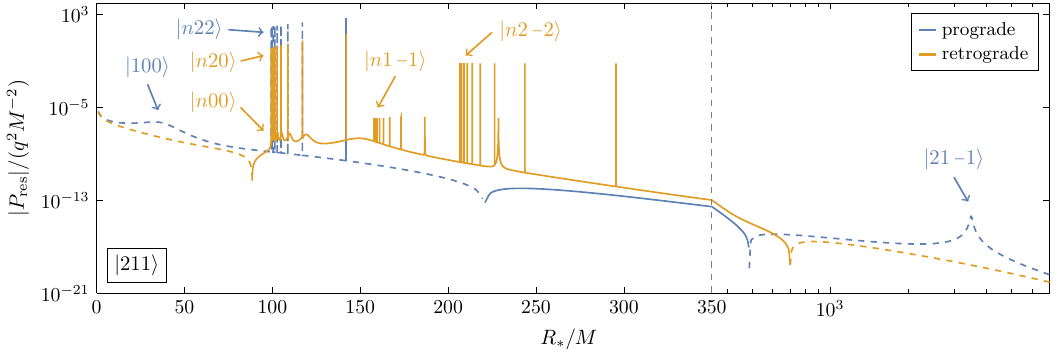}
\caption{Resonance power $P\ped{res}$, defined in \eqref{eqn:Pres}, as a function of the orbital separation $R_*$, for a cloud in the $\ket{211}$ state with $\alpha=0.2$ and $M\ped{c}=0.01M$. We consider here prograde and retrograde circular equatorial orbits, and sum over states with $n\le11$ and $\ell\le2$. The scale of the $x$ axis is linear up to $R_*=350M$, and logarithmic afterwards. We use solid (dashed) lines when $P\ped{res}>0$ ($P\ped{res}<0$), and indicate which states $\ket{b}$ are responsible for each group of resonances visible in the plot.}
\label{fig:Pres211}
\end{figure*}

\begin{figure*}
\centering
\includegraphics[]{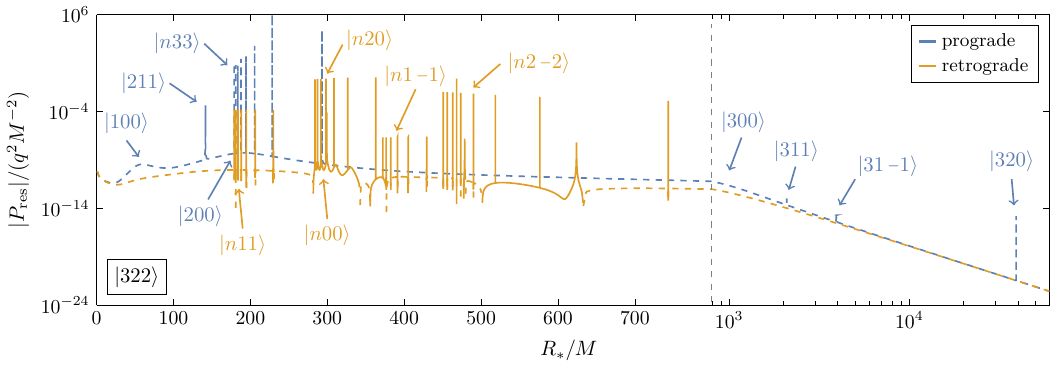}
\caption{Same as Fig.~\ref{fig:Pres211}, but for a cloud in the $\ket{322}$ state. We include states with $n\le11$ and $\ell\le3$ ($\ell\le2$) for the prograde (retrograde) case. Two narrow resonances are present but not easily visible in the prograde plot: with $\ket{21\,\minus1}$ at $R_*=293M$ (close to $\ket{433}$) and with $\ket{32\,\minus2}$ at $R_*=38830M$ (close to $\ket{320}$, but having a much smaller peak).}
\label{fig:Pres322}
\end{figure*}

Before we discuss the range of applicability of \eqref{eqn:Pres}, it is insightful to numerically calculate and plot it. We show $P\ped{res}$ in Figs.~\ref{fig:Pres211} and~\ref{fig:Pres322}, for a cloud in the $\ket{211}$ and $\ket{322}$ states respectively. We consider for the moment equatorial orbits, for which the only nonzero matrix elements have $g=\pm(m-m_a)$, where we suppressed for later convenience the $b$ indices in $\ket{n_b\ell_bm_b}$. The shape of $P\ped{res}$ quickly reveals the positions of the various resonances, which can be grouped into \emph{Bohr}, \emph{fine} and \emph{hyperfine} \cite{Baumann:2018vus,Baumann:2019ztm,Tomaselli:2024bdd}. For a $\ket{211}$ cloud, the only hyperfine resonance on equatorial orbits is with $\ket{21\,\minus1}$, while the only fine resonance is with the $\ket{200}$ state and is too weak to be visible.

The majority of the resonances appear as extremely narrow peaks. This is due to the suppression of the decay rate for nonzero $\ell$, which in the Detweiler approximation reads $\Gamma_b\propto\alpha^{4\ell+5}$ \cite{PhysRevD.22.2323}. States with $\ell=0$ (and sometimes $\ell=1$ for small $n$) constitute an exception, and appear as broad peaks with appreciable width. Due to the large inaccuracies of the Detweiler approximation at moderate values of $\alpha$, we compute the decay rates with the Leaver continued fraction method \cite{Leaver:1985ax,Dolan:2007mj,Baumann:2019eav}, so that the widths and heights of the peaks in the plots are trustworthy.

A clear feature of Figs.~\ref{fig:Pres211} and~\ref{fig:Pres322} is that, at fixed $\ell$ and $m$, the resonances accumulate towards the same point for $n\to\infty$. This happens because $E_b\to0$ for $n\to\infty$, so that the resonant frequency tends to $\Omega=-E_a/g$, independent of $n$ and $\ell$. This observation has far-reaching consequences, which we explore in detail in Sec.~\ref{sec:no-sharp-features}.

We also note that for superradiant states, i.e., those with $\Gamma_b>0$, the sign of $P\ped{res}$ is opposite to that of $E_b-E_a$. Even though in Figs.~\ref{fig:Pres211} and~\ref{fig:Pres322} we do not treat this case differently from the others, it is important to realize that the derivation from Sec.~\ref{sec:fixed-frequency} does not apply for $\Gamma_b>0$. The reason is that the terms $A_-$ and $B_-$ in \eqref{eqn:cat} and \eqref{eqn:cbt} do not decay away, and instead grow exponentially. This instability must however be much slower than all the other timescales of the system, otherwise such a higher-$m$ superradiant mode would have grown, e.g., before the formation of the binary system. This observation naturally leads us to question what happens when the frequency chirp rate $\dot\Omega$ is larger in magnitude than $\Gamma_b$, regardless of the sign of the latter.

\subsection{Chirping frequency}
\label{sec:chirping-frequency}

We derived Eqs.~\eqref{eqn:Mdot} and \eqref{eqn:Pres} under the assumption of a constant frequency $\Omega$. In this Section, we make connection with the case where $\dot\Omega\ne0$, which has been extensively discussed in the literature. As a first step, it is instructive to see what happens when we interpret \eqref{eqn:Mdot} and \eqref{eqn:Pres} as ``instantaneous'' quantities for the case of a slowly evolving $\Omega(t)$. Suppose $\Omega(t)=\gamma t$, and consider for simplicity the case of a two-state system with $\Gamma_b<0$ and $|\eta|^2$ independent of $\Omega$. Then, by integrating \eqref{eqn:Mdot}, we find that the ratio of the final and initial values of $M\ped{c}$ is is given by
\beq
\frac{M\ped{c}(+\infty)}{M\ped{c}(-\infty)}=\exp\biggl(\int_{-\infty}^\infty\frac{\dot M\ped{c}\ap{res}}{M\ped{c}}\dd t\biggr)=e^{-2\pi Z}\,,
\label{eqn:Mcprime}
\eeq
independent of $\Gamma_b$, where $Z\equiv\abs{\eta}^2/(\abs{g}\gamma)$.

This result precisely matches that of a Landau-Zener transition, which describes the evolution of the system \eqref{eqn:schrodinger}, with $\Gamma_a=0$, when the frequency is assumed to increase linearly with time \cite{zener1932non,landau1932theorie,Baumann:2019ztm}. The final occupation of the initial state is indeed known to be independent of $\Gamma_b$, as long as it is negative \cite{PhysRevA.46.4110,PhysRevA.55.2982,Takahashi:2023flk,Boskovic:2024fga}.

We thus see that using formula \eqref{eqn:Mdot} in a scenario where $\dot\Omega\ne0$ yields the correct integrated result for any value of the Landau-Zener parameter $Z$. When $\Gamma_b>0$, one finds instead $M\ped{c}(+\infty)/M\ped{c}(-\infty)=e^{2\pi Z}$, so it is possible to treat the unstable case $\Gamma_b>0$, discussed in Sec.~\ref{sec:numerical-results}, by simply flipping the sign of $\Gamma_b$ before performing the integral.

It should be kept in mind, however, that the instantaneous cloud mass loss rate is only well approximated by \eqref{eqn:Mdot} when $\abs{\Gamma_b}\gg\sqrt\gamma$. In this case, the assumption of slowly evolving frequency is justified, as its chirp rate is much slower than the decay rate of $\ket{b}$, and the resonance width is $\Delta\Omega\sim\abs{\Gamma_b}$. Conversely, when $\abs{\Gamma_b}\ll\sqrt\gamma$, formula \eqref{eqn:Mdot} returns the correct integrated result even though it predicts the incorrect resonance width, which in this case is $\Delta\Omega\sim\sqrt\gamma$ instead. In the general case, the populations $\abs{c_a}^2$ and $\abs{c_b}^2$ of the two states can be written as a function of time in terms of the parabolic cylinder functions.

The reader might wonder, at this point, what is the usefulness of the quantities $\dot M\ped{c}\ap{res}$ and $P\ped{res}$, as defined in \eqref{eqn:Mdot} and \eqref{eqn:Pres}, if they are only valid in a given limit of a well-known general case. The answer is that the time dependence of $\Omega(t)$ is no longer linear when the cloud's backreaction on the orbit is taken into account. This makes the analytical solution useless, while \eqref{eqn:Mdot} and \eqref{eqn:Pres} still apply when the frequency evolves slowly, as we show next.

\section{Backreaction}
\label{sec:backreaction}

\subsection{Brief review of floating resonances}
\label{sec:review-floating}

Resonances between states of the cloud backreact on the orbit, giving rise to \emph{floating} and \emph{sinking} orbits, depending on the sign of $E_a-E_b$ \cite{Baumann:2019ztm}, as well as by altering the orbital inclination and eccentricity \cite{Tomaselli:2024bdd,Boskovic:2024fga,Tomaselli:2024dbw}. We focus here on circular equatorial orbits for simplicity. When we require the total energy of the binary-cloud system to be conserved, we find that the frequency increases according to \cite{Tomaselli:2024bdd}
\beq
\dot\Omega=\gamma+\tilde B\frac{\dd\abs{c_a}^2}{\dd t}\,.
\label{eqn:dOmegadt}
\eeq
In this equation, $\gamma$ is the chirp rate induced by external energy losses (such as GW radiation or dynamical friction): if $P>0$ is the power lost, then to leading order in $q$ we have
\beq
\gamma=\frac{3\Omega^{1/3}}{qM^{5/3}}P\,.
\eeq
The parameter
\beq
\tilde B=\frac{3M\ped{c}(\Omega M)^{1/3}(E_a-E_b)}{Mq\alpha}\,,
\eeq
where $M\ped{c}$ is the initial value of the cloud's mass, quantifies the magnitude of the backreaction due to the resonance.\footnote{The parameter $\tilde B$ is related to the dimensionless backreaction parameter $B$ defined in \cite{Tomaselli:2024bdd} by $B\sqrt{\gamma/\abs{g}}=\tilde B\big|_{g\Omega=E_b-E_a}$.}

\begin{figure}
\centering
\includegraphics[]{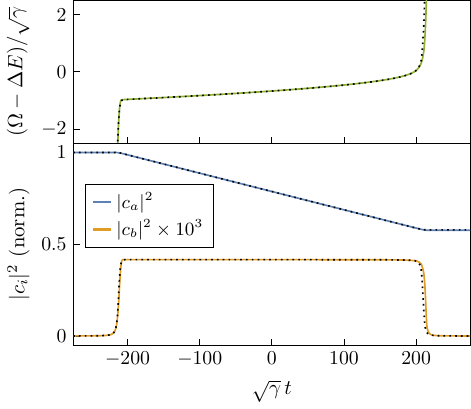}
\caption{Example of a floating resonance with $Z=10^{-3}$, $\tilde B=10^3\sqrt\gamma$, $\Gamma_b=-1.2\sqrt\gamma$, and $g=-1$. We normalize $c_a$ and $c_b$ such that $\abs{c_a}^2+\abs{c_b}^2=1$ at $t=-\infty$, and define $\Delta E=E_b-E_a$. We obtain the solid colored lines by solving the Schrödinger equation \eqref{eqn:schrodinger} coupled to Eq.~\eqref{eqn:dOmegadt}, and the black dotted lines by solving Eqs.~\eqref{eqn:P+Pres} and \eqref{eqn:McMcres}, with \eqref{eqn:b-amplitude} used to calculate $\abs{c_b}^2$. This effective description of the resonance is in excellent agreement with the solution of the nonlinear Schrödinger equation.}
\label{fig:Gamma-breaking}
\end{figure}

The phenomenology of the nonlinear system constituted by the coupled Eqs.~\eqref{eqn:schrodinger} and \eqref{eqn:dOmegadt} has been studied in detail in \cite{Tomaselli:2024bdd}. We show in Fig.~\ref{fig:Gamma-breaking} an example of a floating orbit, obtained by solving numerically the Schrödinger equation \eqref{eqn:schrodinger} coupled to Eq.~\eqref{eqn:dOmegadt}. The population of state $\ket{a}$ steadily decreases during the resonance, while the frequency ``floats'' at an approximately constant value. This behavior continues until $\abs{c_a}^2$ reaches a certain critical threshold, when the resonance suddenly \emph{breaks}. The corresponding value of the cloud's mass was determined analytically in \cite{Tomaselli:2024bdd} by studying the nonlinear Schrödinger equation, and equals
\beq
\frac{M\ped{c}\ap{break}}{M\ped{c}}=\frac{-\Gamma_b}{2Z\tilde B}\,.
\label{eqn:Mcbreak}
\eeq
When formula \eqref{eqn:Mcbreak} returns a value larger than 1, no float is observed in the frequency evolution.

\subsection{Effective description of resonance backreaction}
\label{sec:effective-resonances}

We now show that we can recover the physics described in Sec.~\ref{sec:review-floating} using $\dot M\ped{c}\ap{res}$ and $P\ped{res}$, defined in \eqref{eqn:Mdot} and \eqref{eqn:Pres}. This provides new physical insight and eliminates the need to solve the Schrödinger equation numerically. Suppose then that the resonance induces an energy loss $P\ped{res}$ on the binary, such that the frequency evolves according to
\beq
\dot\Omega=\frac{3\Omega^{1/3}}{qM^{5/3}}\bigl(P+P\ped{res}\bigr)\,.
\label{eqn:P+Pres}
\eeq
Note that $P\ped{res}<0$ for $E_a>E_b$, indicating that the binary gains energy from the resonance, as expected on a floating orbit. The power $P\ped{res}$ is proportional to the instantaneous value of the cloud's mass $M\ped{c}$, which itself evolves as
\beq
\dot M\ped{c}=\dot M\ped{c}\ap{res}\,.
\label{eqn:McMcres}
\eeq

Equations \eqref{eqn:P+Pres} and \eqref{eqn:McMcres} offer a simple interpretation of the mechanism regulating floating orbits and their breaking. Because $P\ped{res}$ is negative and peaked around $g\Omega=E_b-E_a$, at low enough frequencies we have $P\gg\abs{P\ped{res}}$ and thus $\dot\Omega>0$. Conversely, close enough to the peak we have $\abs{P\ped{res}\ap{peak}}\gg P$ and thus $\dot\Omega<0$. There exists therefore a stable fixed point for the frequency evolution, where $P=-P\ped{res}$ and $\dot\Omega=0$, located at a frequency just below the resonant value $(E_b-E_a)/g$. While the frequency floats, sitting at this stable fixed point, the cloud's mass slowly decreases according to \eqref{eqn:McMcres}. This moves the frequency fixed point closer and closer to $(E_b-E_a)/g$, as can be seen from the slow upward drift of $\Omega$ in Fig.~\ref{fig:Gamma-breaking}. When the cloud has been depleted so much that $P>\abs{P\ped{res}\ap{peak}}$, the fixed point disappears and we have $\dot\Omega>0$ for any $\Omega$: this is the point where the resonance breaks. Quantitatively, by inserting $g\Omega=E_b-E_a$ in \eqref{eqn:Pres}, we have
\beq
P\ped{res}\ap{peak}=\frac{M\ped{c}}\mu\frac{2(E_a-E_b)\abs{\eta}^2}{\Gamma_b}\,.
\eeq
Imposing that $P\ped{res}\ap{peak}$ equals $-P$, we find the cloud's mass at resonance breaking,
\beq
M\ped{c}\ap{break}=\frac{-\Gamma_b\mu P}{2(E_a-E_b)\abs{\eta}^2}\,,
\eeq
which, after substituting the definitions of $\gamma$, $Z$ and $\tilde B$, agrees exactly with \eqref{eqn:Mcbreak}. This approach also clarifies why no float is observed when $M\ped{c}\ap{break}>M\ped{c}$: there is no stable fixed point to begin with.

This effective description of resonances not only gives the correct resonance-breaking condition, but the numerical solution of \eqref{eqn:P+Pres} and \eqref{eqn:McMcres} is in excellent agreement with that of \eqref{eqn:schrodinger} and \eqref{eqn:dOmegadt} throughout the whole resonance, as we show in Fig.~\ref{fig:Gamma-breaking}.

The advantage offered by this approach might seem marginal, when one can also directly solve the Schrödinger equation. However, the latter method is only practical when the Hamiltonian is expressed in a dressed frame, as in \eqref{eqn:H_D}. A dressed frame transformation is not available, e.g., when the coupling $\braket{a|V_*|b}$ includes multiple Fourier components, and to deal with their fast oscillations one would need to choose an extremely small numerical time step. This situation arises when considering orbits with generic inclination and eccentricity, as we discuss in Sec.~\ref{sec:eccentric-inclined}.

Furthermore, the treatment of resonances in previous works \cite{Baumann:2019ztm,Tomaselli:2024bdd} relied on neglecting the frequency dependence of key quantities such as $\abs{\eta}^2$ and $\tilde B$. While this approximation is useful to study the Schrödinger equation both analytically and numerically, the shape of $P\ped{res}$ demonstrates that resonances with states $\ket{n00}$ occur over a very extended range of frequencies, due to their large decay width $\abs{\Gamma_b}$. We therefore expect the method described here to be more accurate in those cases, compared to results previously found in the literature.

As anticipated in Sec.~\ref{sec:chirping-frequency}, for narrow resonances with $\abs{\Gamma_b}<\sqrt{\gamma}$,\footnote{When backreaction is included, this condition is upgraded to $\abs{\Gamma_b}<\dot\Omega^{1/2}$. When backreaction is strong, the dependence of $\Omega$ on $\Gamma_b$ makes checking this condition not always trivial.} the time evolution is not correctly predicted by $\dot M\ped{c}\ap{res}$ and $P\ped{res}$, and one needs to solve the Schrödinger equation. The applicability of this effective description of resonances largely depends, therefore, on the magnitude of the external energy losses. Beyond GW radiation, these include the cloud's ionization, which we explore in Sec.~\ref{sec:no-sharp-features}.

\section{Wide resonances phenomenology}
\label{sec:wide-resonance-phenomenology}

\subsection{Inclined and eccentric orbits}
\label{sec:eccentric-inclined}

\begin{figure*}
\centering
\includegraphics[]{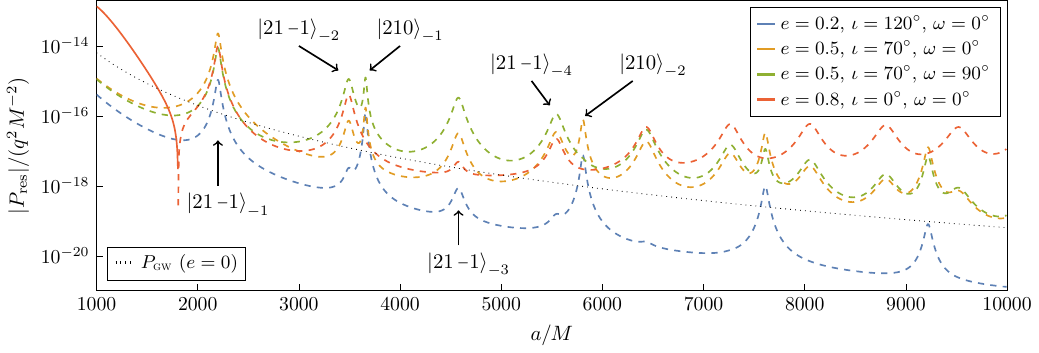}
\caption{Resonance power as a function of the semi-major axis $a$, for orbits with generic eccentricity $e$, inclination $\iota$, and argument of periapsis $\omega$, for a cloud in the $\ket{211}$ state with $\alpha=0.2$ and $M\ped{c}=0.01M$. We sum here over $n\le3$, $\ell\le1$, and $\abs{g}\le11$. As in Fig.~\ref{fig:Pres211}, we use solid (dashed) lines when $P\ped{res}>0$ ($P\ped{res}<0$) and label the various peaks with the state responsible for the resonance, using the notation $\ket{n\ell m}_g$. The thin dotted line shows the radiated GW $P_\slab{gw}$ for a circular orbit.}
\label{fig:Pres211_generic_orbit}
\end{figure*}

The results shown in the previous sections can be extended to orbits with generic eccentricity $e$, inclination with respect to the equatorial plane $\iota$, and argument of periapsis $\omega$. Note that $\iota=\SI{0}{\degree}$ ($\SI{180}{\degree}$) corresponds to the prograde (retrograde) case discussed earlier, and the parameter $\omega$ is only needed when both $e$ and $\iota$ are nonzero. While on equatorial circular orbits the only nonzero matrix elements have $g=\pm(m-m_a)$, on generic orbits any integer value of $g$ can in principle contribute, creating its own resonance. The phenomenology of eccentric and/or inclined orbits has previously been studied in \cite{Berti:2019wnn,Tomaselli:2023ysb,Fan:2023jjj,Tomaselli:2024bdd,Tomaselli:2024dbw,Boskovic:2024fga}, but some aspects of it can be more effectively understood by simply computing $\dot M\ped{c}\ap{res}$ and $P\ped{res}$.

We show in Fig.~\ref{fig:Pres211_generic_orbit} the resonance power $P\ped{res}$, for $\ket{a}=\ket{211}$, on a few selected orbits. We focus here on hyperfine resonances, because the Bohr region becomes so extremely crowded with resonances that it is hard to extract useful information from it. We see that mildly eccentric, highly inclined orbits ($e=0.2$, $\iota=\SI{120}{\degree}$) predominantly excite overtones of the resonance with $\ket{210}$. Conversely, highly eccentric equatorial orbits ($e=0.8$, $\iota=\SI{0}{\degree}$) strongly excite all overtones of the resonance $\ket{21\,\minus1}$, while not exciting any with $\ket{210}$. Varying the argument of periapsis between $\SI{0}{\degree}$ and $\SI{90}{\degree}$ slightly changes the magnitude of $P\ped{res}$ and the relative amplitudes of its peaks, but does not turn off any resonance, at least for orbits with intermediate values of eccentricity and inclination ($e=0.5$, $\iota=\SI{70}{\degree}$).

It is interesting to see that all resonances appear to be clearly separated from each other.\footnote{A similar conclusion was claimed in \cite{Tomaselli:2024bdd}, based however on an incorrect argument, where the resonance width was assumed to be of order $\Delta\Omega\sim\eta$. As discussed at length in this paper, for wide resonances the width is instead set by $\abs{\Gamma_b}$.} This is not \emph{a priori} guaranteed: because to leading order $E_{211}-E_{21\,\minus1}\approx2(E_{211}-E_{210})$, one could expect resonances with $\ket{21\,\minus1}$, with a given value of $g$, to be degenerate with those with $\ket{210}$ and half that value of $g$. We are able to break this degeneracy because we compute both the bound state energies and the resonance widths with Leaver's method.

The ``resonant history'' of gravitational atoms in binary systems was explored systematically in \cite{Tomaselli:2024bdd}. This work found that the fate of the binary system and the cloud is highly dependent on the ``strongest'' hyperfine resonance encountered during the evolution, especially its breaking point as determined by \eqref{eqn:Mcbreak}. We now see that this analysis can qualitatively be performed almost by eye, comparing $P\ped{res}$ on generic orbits with the power lost to GW radiation $P_\slab{gw}$ (or other types of external energy losses, if applicable). We leave a re-analysis of the resonant history with the methods developed here for a future work.

\subsection{The $\ket{100}$ floating orbit}

A very interesting application of our effective description of wide backreacted resonances is the $\ket{211}\to\ket{100}$ floating orbit with $g=-1$. This resonance, clearly visible as a broad peak in Fig.~\ref{fig:Pres211}, is the last one encountered by a binary system before merger. Due to its atypically large width $\abs{\Gamma_b}$, it was left out of the systematic study of resonances performed in \cite{Tomaselli:2024bdd}.

\begin{figure}
\centering
\includegraphics[]{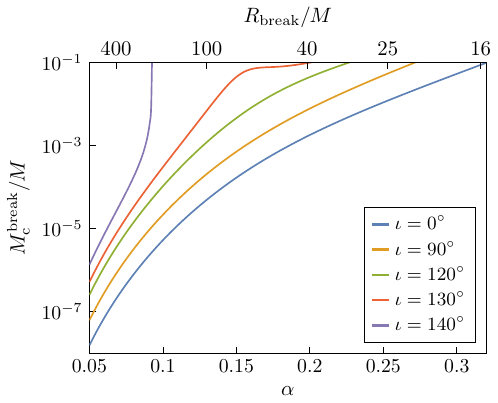}
\caption{Cloud's mass at the breaking of the $\ket{211}\to\ket{100}$ resonance, as a function of $\alpha$, for circular orbits with various inclinations $\iota$. The binary is stuck  in a floating orbit at $R_*\gtrsim R\ped{break}$ (shown at the top of the $x$ axis) until the cloud's mass drops below $M\ped{c}\ap{break}$, at which point the binary can proceed to merger. In a wide region of parameter space, the cloud's mass that can survive until merger is significantly limited by this last resonance.}
\label{fig:Mc_break_100}
\end{figure}

A floating orbit is expected whenever the total power lost,
\beq
P_\slab{gw}+P\ped{ion}+P\ped{res}\,,
\eeq
where $P_\slab{gw}=(32/5)q^2M^5/R^5$ is the radiated GW power and $P\ped{ion}$ is the ionization contribution (discussed in Sec.~\ref{sec:no-sharp-features}), goes from positive to negative as the frequency increases. By using the results of \cite{Tomaselli:2023ysb} for $P\ped{ion}$ on inclined orbits, we can thus easily check for which values of the parameters there exists a value of $\Omega$ where this occurs. The mass of the cloud $M\ped{c}$ controls the amplitude of $P\ped{ion}$ and $P\ped{res}$. As $M\ped{c}$ decreases, we expect $P\ped{ion}$ and $P\ped{res}$ to become subdominant compared to $P_\slab{gw}$, and the floating orbit disappears when $M\ped{c}=M\ped{c}\ap{break}$, at $R_*=R\ped{break}$. Furthermore, as the orbital inclination $\iota$ is varied from $\SI{0}{\degree}$ (prograde) to $\SI{180}{\degree}$ (retrograde), the resonance weakens and eventually vanishes, like any other floating orbit.

We show in Fig.~\ref{fig:Mc_break_100} the minimum value of $M\ped{c}$ for which the $\ket{211}\to\ket{100}$ floating orbit occurs, as a function of $\alpha$ and for different values of the inclination $\iota$. As expected, we find that $M\ped{c}\ap{break}$ increases with $\iota$, and the floating orbit is not observed at all for $\iota\gtrsim\SI{140}{\degree}$ and $\alpha\gtrsim0.09$. The small value of $M\ped{c}\ap{break}$ seen in other cases is particularly consequential for GW observations, as it provides an upper bound for the cloud's mass $M\ped{c}$ that can survive up to the last stages of the inspiral before merger. This is where the binary's GW signal is loudest, and it is therefore the easiest place to look for the cloud's signature, at least from the data analysis point of view. The actual mass of the cloud at merger will be smaller than $M\ped{c}\ap{break}$, as ionization and the off-resonance flux into $\ket{100}$ keep depleting the cloud even after the floating orbit breaks. Our discussion also implies that it is never possible for $P\ped{ion}$ or $P\ped{res}$ to dominate over $P_\slab{gw}$ in the region $R_*<R\ped{break}$, at least for systems with small enough orbital inclinations.

The constraint derived here applies predominantly for $\iota\lesssim\SI{140}{\degree}$, where hyperfine resonances are also expected to destroy the cloud, at a much earlier stage of the inspiral. However, binaries can ``skip'' hyperfine resonances depending on their formation mechanism \cite{Tomaselli:2024bdd}, in which case the constraint in Fig.~\ref{fig:Mc_break_100} becomes relevant.

Floating orbits are also known to significantly modify the orbital inclination and eccentricity \cite{Tomaselli:2024bdd,Boskovic:2024fga,Tomaselli:2024dbw}, for example bringing the binary close to a nonzero eccentricity fixed point. The $\ket{100}$ resonance would be particularly exciting in this regard, as it happens much closer to merger compared to the other (hyperfine) floating resonances, implying that GW radiation has much less time to circularize the binary before merger. Unfortunately, however, the eccentricity fixed point of resonances with $(m-m_a)/g=1$, like the one we are considering, is zero. It is still possible that the cloud gets destroyed at an earlier instance of the $\ket{211}\to\ket{100}$ resonance, with a higher value of $g$, which could thus excite the eccentricity. We leave this question for a future study.

\section{No sharp features}
\label{sec:no-sharp-features}

\subsection{Brief review of ionization}
\label{sec:review-ionization}

Apart from the resonances described earlier, the perturbation from the binary companion also excites unbound states \cite{Baumann:2021fkf,Baumann:2022pkl}. This process is known as \emph{ionization} and it backreacts on the orbit by inducing dynamical friction \cite{Tomaselli:2023ysb} on the companion.

We label unbound states with their wavenumber $k$ and the usual angular momentum numbers $\ell$ and $m$, and normalize them according to $\braket{k'\ell'm'|k\ell m}=2\pi\delta_{\ell'\ell}\delta_{m'm}\delta(k'-k)$. When the orbital frequency $\Omega$ is kept fixed, the cloud's mass is ionized at a rate
\beq
\dot M\ped{c}\ap{ion}=M\ped{c}\sum_{\ell,g}\frac{\mu\abs{\tilde\eta}^2}{k_g}\Theta\bigl(k_g^2\bigr)\,,
\label{eqn:Mcion}
\eeq
where $k_g=\sqrt{2\mu(E_a+g\Omega)}$ and $\tilde\eta=\braket{a|V_*|k_g\ell m}$. Note that, while $\eta$ has dimensions of an energy, due to the $\delta$-function normalization of unbound states $\tilde\eta$ has dimensions of an energy times square root length. The \emph{ionization power} lost by the binary in the process is instead given by
\beq
P\ped{ion}=\frac{M\ped{c}}\mu\sum_{\ell,g}g\Omega\frac{\mu\abs{\tilde\eta}^2}{k_g}\Theta\bigl(k_g^2\bigr)\,.
\label{eqn:Pion}
\eeq

The $\Theta$ functions appearing in \eqref{eqn:Mcion} and \eqref{eqn:Pion} arise because there is a minimum energy gap separating the cloud's state $\ket{a}$, with energy $E_a<0$, from any given unbound state, with positive energy $k^2/(2\mu)>0$. Only perturbations with frequency above the threshold
\beq
\Omega_{(g)}=-\frac{E_a}g\,,
\label{eqn:Omega_g}
\eeq
for a given Fourier mode $g$, can therefore ionize the cloud. These frequencies also correspond to a finite jump of $\dot M\ped{c}\ap{ion}$ and $P\ped{ion}$. Indeed, the long-range gravitational potential of the BH confines the wavefunctions of marginally bound states just enough to make the following limit nonzero,
\beq
\tilde\xi\equiv\lim_{k\to0}\frac{\abs{\tilde\eta}^2}k\,,
\eeq
as discussed, e.g., in Appendix D of \cite{Baumann:2021fkf}. The existence of such jumps has been confirmed by fully relativistic calculations \cite{Brito:2023pyl,Dyson:2025dlj,Li:2025ffh}.

We plot $P\ped{ion}$ in Fig.~\ref{fig:sharp-features}, showing its first two discontinuities, at $\Omega_{(1)}$ and $\Omega_{(2)}$. A nonzero frequency chirp $\dot\Omega=\gamma$ has the effect of smoothening these jumps, over a frequency range $\Delta\Omega\sim\sqrt\gamma$ \cite{Baumann:2021fkf}. This is analogous to the case of resonances narrower than $\sqrt\gamma$ we discussed in Sec.~\ref{sec:chirping-frequency}.

When $P\ped{ion}$ is used to determine the time evolution of the system through a conservation equation, such as
\beq
\dot\Omega=\frac{3\Omega^{1/3}}{qM^{5/3}}(P_\slab{gw}+P\ped{ion})\,,
\label{eqn:omegadot-pgw-pion}
\eeq
its jumps create \emph{kinks} in the frequency, i.e., discontinuities in its first derivative, as we show in Fig.~\ref{fig:kinks}. Similarly, the jumps of $\dot M\ped{c}\ap{ion}$ create kinks in the cloud's mass as a function of time.

These frequency kinks, imprinted on the emitted gravitational waveform, have been claimed to be a smoking gun of the interaction with the cloud \cite{Baumann:2022pkl}. We now show that this conclusion ceases to be true when the backreaction of resonances is correctly taken into account.

\subsection{Accumulation of resonances}
\label{sec:accumulation}

\begin{figure}
\centering
\includegraphics[]{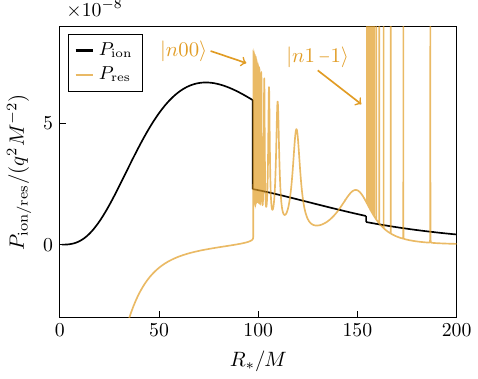}
\caption{Comparison between $P\ped{ion}$ and $P\ped{res}$ on a retrograde orbit. We use the same parameters as in Fig.~\ref{fig:Pres211}, but include states with $n\le30$ and $\ell\le1$ in the computation of $P\ped{res}$. Differently from previous works, we use bound state energies determined with Leaver's method (rather their nonrelativistic approximation) in the computation of $P\ped{ion}$. This way, the positions of the ionization jumps match that the accumulation points of the resonances seen in $P\ped{res}$.}
\label{fig:sharp-features}
\end{figure}

As we noticed in Sec.~\ref{sec:numerical-results}, resonances with states $\ket{n\ell m}$ accumulate towards a common point for $n\to\infty$, when $\ell$ and $m$ are kept fixed. Because $E_b\to0$ for $n\to\infty$, this point corresponds to the ionization threshold frequency $\Omega_{(g)}$ in \eqref{eqn:Omega_g}, as was pointed out in previous works \cite{Brito:2023pyl,Tomaselli:2024bdd,Tomaselli:2024dbw}. This property is particularly evident when we show $P\ped{ion}$ and $P\ped{res}$ side by side, as in Fig.~\ref{fig:sharp-features}. We now calculate the combined effect of these infinitely many closely spaced resonances, and show that it exactly cancels the ionization jump.

The cloud's bound state spectrum becomes hydrogenic not only for $\alpha\to0$, but also for $n\to\infty$. This is because the higher the energy of a bound state, the more spread out its wavefunction, implying that the Newtonian approximation becomes increasingly better. The energy difference between bound states with consecutive values of $n$ is thus
\beq
\Delta E_{n}=-\frac{\mu\alpha^2}{2(n+1)^2}+\frac{\mu\alpha^2}{2n^2}\approx\frac{\mu\alpha^2}{n^3}\,,
\eeq
for $n\to\infty$. Their wavefunctions converge to
\beq
\psi(\vec r)\xrightarrow{n\to\infty}\frac{2(\mu\alpha)^{3/2}}{n^{3/2}\sqrt{2\mu\alpha r}}\,J_{2\ell+1}(2\sqrt{2\mu\alpha r})Y_{\ell m}(\hat{\vec r})
\eeq
where $J_{2\ell+1}$ is a Bessel function of the first kind and the normalization $\sim n^{-3/2}$ arises from the increasing spatial extent of the wavefunction. We thus have $\abs{\eta}^2\sim n^{-3}$, and the limit
\beq
\xi\equiv\lim_{n\to\infty}\frac{\abs{\eta}^2}{\Delta E_{n}}
\label{eqn:xi}
\eeq
is finite.

Let us now use the approach of Sec.~\ref{sec:chirping-frequency}, where we apply the fixed-frequency results, such as $\dot M\ped{c}\ap{res}$, with a slowly evolving $\Omega=\gamma t$. Each resonance reduces the cloud's mass by an amount
\beq
\Delta M\ped{c}\ap{res}=\frac{2\pi M\ped{c}\abs{\eta}^2}{g\gamma}\,,
\label{eqn:DeltaMc}
\eeq
according to the small-$Z$ limit of \eqref{eqn:Mcprime}. We can then define the time-averaged cloud's mass loss in the large-$n$ limit as
\beq
\braket{\dot M\ped{c}\ap{res}}=\lim_{n\to\infty}\frac{\Delta M\ped{c}\ap{res}}{\Delta\Omega_n}\times\frac{\Delta\Omega_n}{\Delta t_n}=2\pi M\ped{c}\xi\,,
\label{eqn:averageMcdot}
\eeq
where $\Delta\Omega_n=\Delta E_n/g$ is the difference of resonant frequencies of two consecutive resonances, and $\Delta t_n=\Delta\Omega_n/\gamma$. The result \eqref{eqn:averageMcdot} must then be summed over $\ell$ and $m$ to obtain the total contribution from all resonances. We thus see that, at least after taking a suitable average, bound states also introduce a discontinuity at $\Omega_{(g)}$. However, $\braket{\dot M\ped{c}\ap{res}}$ jumps \emph{down} as $\Omega$ increases, while $\dot M\ped{c}\ap{ion}$ jumps \emph{up}.

We now show that the amplitudes of the jumps are identical. The key insight is that the zero-energy mode can be obtained both as a high-energy limit of bound states, or a low-energy limit of unbound states. The limit \eqref{eqn:xi} must then also give the same result when taken from the positive-energy side. However, Eq.~\eqref{eqn:xi} involves the overlap $\abs{\eta}^2$ of $V_*$ between bound states, while $\dot M\ped{c}\ap{ion}$ involves the \emph{density} of overlaps $\abs{\tilde\eta}^2$. We can relate the two quantities if we imagine to discretize the continuum into small intervals with wavenumbers $[k,k+\Delta k]$. We can then identify the overlap $\abs{\eta}^2$ with the integrated continuum overlap of the interval,
\beq
\abs{\eta}^2\overset{!}{=}\int_k^{k+\Delta k}\abs{\tilde\eta}^2\frac{\dd k}{2\pi}=\abs{\tilde\eta}^2\frac{\Delta k}{2\pi}\,,
\eeq
and accordingly replace $\Delta E_n$ with the energy difference of the unbound states at the edge of the interval. We then obtain
\beq
\xi=\lim_{k\to0}\lim_{\Delta k\to0}\frac{\abs{\tilde\eta}^2\Delta k/(2\pi)}{\frac{(k+\Delta k)^2}{2\mu}-\frac{k^2}{2\mu}}=\lim_{k\to0}\frac{\mu\abs{\tilde\eta}^2}{2\pi k}=\frac{\mu}{2\pi}\tilde\xi\,.
\label{eqn:xi-xitilde}
\eeq
Alternatively, and perhaps more explicitly, the relation between $\xi$ and $\tilde\xi$ can be found directly from the zero-energy limit of unbound states wavefunctions \cite{Baumann:2021fkf},
\beq
\psi(\vec r)\xrightarrow{k\to0}\sqrt{\frac{4\pi k}r}\,J_{2\ell+1}(2\sqrt{2\mu\alpha r})Y_{\ell m}(\hat{\vec r})\,.
\eeq
After substitution in \eqref{eqn:averageMcdot}, formula \eqref{eqn:xi-xitilde} proves that the discontinuities of $\dot M\ped{c}\ap{ion}$ and $\dot M\ped{c}\ap{res}$  have the same amplitude. The same argument holds for $P\ped{ion}$ and $P\ped{res}$.

\begin{figure}
\centering
\includegraphics[]{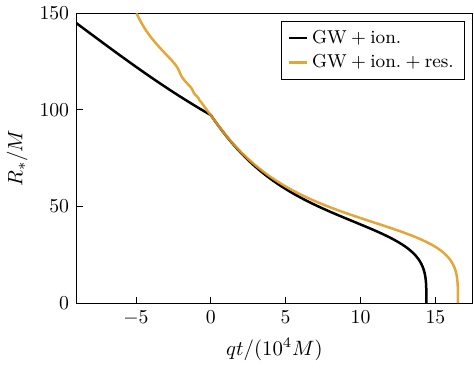}
\caption{Evolution of the separation of a retrograde binary system, under the combined effects of GW radiation and ionization [{\color{black}\textbf{black}}]. A kink is observed at $R_*\approx97M$, corresponding to the first discontinuity of $P\ped{ion}$ in Fig.~\ref{fig:sharp-features}, at the threshold frequency $\Omega_{(1)}$. The $x$ axis has been shifted so that the kink happens at $t=0$. When resonances are included, by adding $M\ped{c}\ap{res}$ to $M\ped{c}\ap{ion}$ and $P\ped{res}$ to $P\ped{ion}$, the kink disappears [{\color{Mathematica2}\textbf{orange}}].}
\label{fig:kinks}
\end{figure}

When solving for the time evolution of the system, such as through \eqref{eqn:omegadot-pgw-pion} with $P\ped{ion}$ replaced by $P\ped{ion}+P\ped{res}$, the time average defined in \eqref{eqn:averageMcdot} determines the slope of integrated quantities in the limit where $\Omega$ approaches $\Omega_{(g)}$. We show in Fig.~\ref{fig:kinks} that the inclusion of $P\ped{res}$ completely eliminates the ionization kink.\footnote{Strictly speaking, the data for $P\ped{res}$ plotted in Fig.~\ref{fig:sharp-features} and used in Fig.~\ref{fig:kinks} only include states with $\ell\le1$. However, all states of the form $\ket{n\ell0}$, with $\ell$ even, create resonances that accumulate at $\Omega_{(1)}$, such as the states $\ket{n20}$ seen in Fig.~\ref{fig:Pres211}. The $\ell=0$ contribution is dominant, hence why the discontinuity in Fig.~\ref{fig:kinks} appears to be eliminated when resonances are included. There is still however a very small residual jump in slope, which would only be removed by the inclusion of higher-$\ell$ states.} While the example shown in Fig.~\ref{fig:kinks} involves resonances with states $\ket{n00}$, which have appreciable width, in general most of the accumulating resonances will be extremely narrow, as seen in Figs.~\ref{fig:Pres211} and~\ref{fig:Pres322}.\footnote{Furthermore, as a consequence of the discussion in Sec.~\ref{sec:chirping-frequency}, the sign of $\Gamma_b$ for superradiant states must be flipped to obtain the correct sign of $\Delta M\ped{c}$.} It is therefore numerically challenging to properly resolve these features when integrating in time. Undersampling $P\ped{res}$ will lead to missing most of the resonance contribution, and the ionization discontinuity will not be cured. To the best of our knowledge, Ref.~\cite{DellaMonica:2025zby} is the only work that, so far, modeled the binary time evolution including a quantity similar to $P\ped{res}$---albeit using a slightly different formula, as we mentioned after Eq.~\eqref{eqn:Pres}. The authors claim to see the ionization-induced kinks, which we speculate to be a consequence of the aforementioned numerical subtleties.

We have so far worked with the fixed-frequency quantities $\dot M\ped{c}\ap{res}$ and $P\ped{res}$, assuming a slow evolution $\Omega=\gamma t$. As discussed in Sec.~\ref{sec:chirping-frequency}, however, this approach is only justified as long as $\abs{\Gamma_b}\gg\sqrt\gamma$. But in the $n\to\infty$ limit, from Detweiler's formula we have $\abs{\Gamma_b}\sim n^{-3}$, so the derivation above must break down for large enough $n$. Luckily, while each resonance gets broadened to a width $\Delta\Omega\sim\sqrt\gamma$, Eq.~\eqref{eqn:DeltaMc} remains unchanged for $n\to\infty$, due to $\abs{\eta}^2\sim n^{-3}\to0$. We then find that $\dot M\ped{c}\ap{res}$ and $P\ped{res}$ will still feature a jump with the same amplitude as before, but smoothened over an interval $\Delta\Omega\sim\sqrt\gamma$. As mentioned in Sec.~\ref{sec:review-ionization}, the jumps of $\dot M\ped{c}\ap{ion}$ and $P\ped{ion}$ also get smoothened, exactly in the same way. We conclude that, when the frequency has a nonzero chip rate, the sharp features are eliminated not only in a time-averaged sense, but also at the level of instantaneous quantities, removing the oscillations seen in Fig.~\ref{fig:sharp-features}.

The reader might wonder whether the resonance backreaction could change their behavior enough to spoil their large-$n$ properties discussed here. The answer is negative: as $Z\to0$, the conditions for strong backreaction derived in \cite{Tomaselli:2024bdd} fail to be satisfied both for floating and sinking resonances.

The absence of sharp features has physical reasons. Unlike some examples of many-body systems encountered in condensed matter, the gravitational atom does not have a gapped energy spectrum, and the bound states smoothly blend into the continuum. We can make this idea concrete if we imagine to place the system in a large, but finite-volume, box. The boundary conditions at the box walls discretize the continuum states into a set of closely spaced bound states, so that no threshold frequency, and thus no ionization jump, exists. But in the limit of infinite volume, we must recover the usual physics of the gravitational atom, which cannot therefore feature any such sharp feature. The same considerations apply to other systems, such as boson stars \cite{Duque:2023seg}, for which ionization-like discontinuities have been claimed.

\section{Comparison with relativistic results}
\label{sec:comparison}

Research on the phenomenology of boson clouds in binary systems has recently been enriched with fully relativistic treatments. The approach was pioneered in \cite{Brito:2023pyl}, which computed in a Schwarzschild background the scalar flux to infinity and through the BH horizon, for both prograde and retrograde equatorial circular orbits. The results have then been extended to Kerr black holes in \cite{Dyson:2025dlj}, although only for prograde orbits, and a new semi-analytical approach has been proposed in \cite{Li:2025ffh}. These works quantitatively agree with each other.\footnote{Although \cite{Dyson:2025dlj} initially claimed a discrepancy with \cite{Brito:2023pyl}, this was later found to be caused by a minor mistake in \cite{Brito:2023pyl}.} Other perturbative formalisms to attack the problem have also been proposed, see e.g.\ \cite{Cannizzaro:2023jle}.

\begin{figure*}
\centering
\includegraphics[]{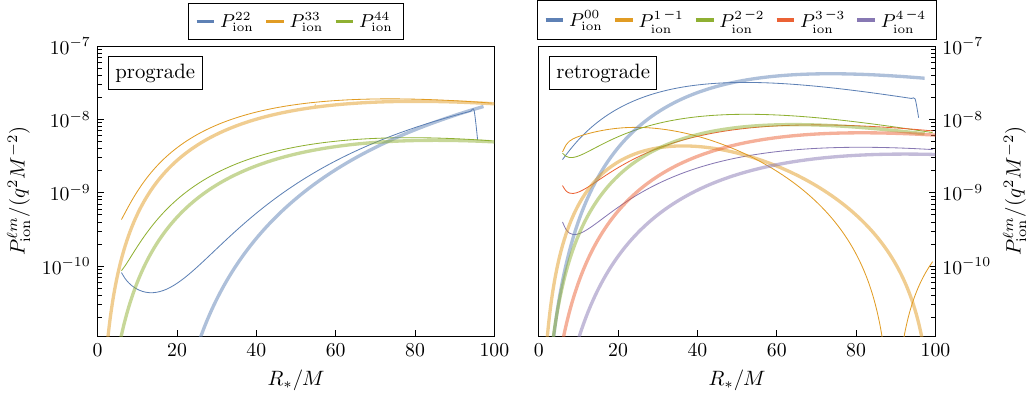}
\caption{Comparison between selected $(\ell,m)$ angular components of $P\ped{ion}$ (thick lines) and the relativistic scalar infinity flux of~\cite{Brito:2023pyl} (thin lines). The latter agrees with other relativistic works \cite{Dyson:2025dlj,Li:2025ffh}. Note that $P\ped{ion}^{00}$ vanishes for $R_*\gtrsim97M$, corresponding to the first ionization discontinuity (the relativistic flux is seen decreasing just before this point due to finite numerical resolution). We assume $\alpha=0.2$ and $M\ped{c}=0.01M$.}
\label{fig:Pion-comparison}
\end{figure*}

\begin{figure*}
\centering
\includegraphics[]{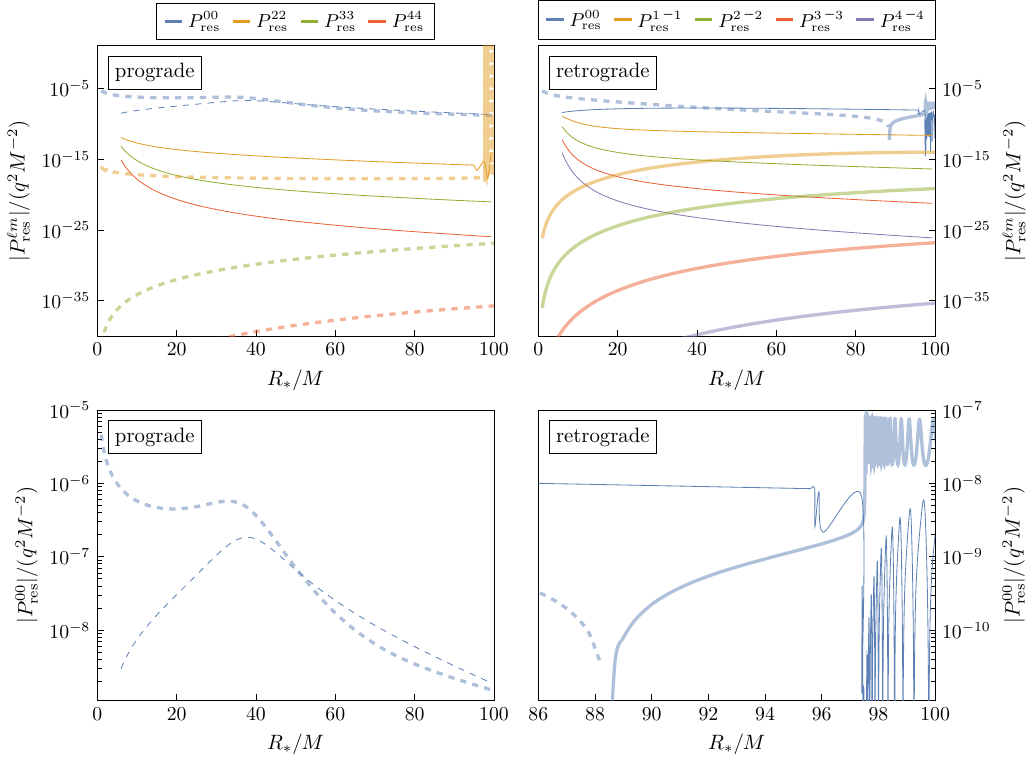}
\caption{Top panels: same as Fig.~\ref{fig:Pion-comparison}, but comparing $P\ped{res}$ (thick lines) and the horizon flux of \cite{Brito:2023pyl,Dyson:2025dlj,Li:2025ffh} (thin lines). Bottom panels: detail on the comparison of the dominant $(\ell,m)=(0,0)$ modes. As in Fig.~\ref{fig:Pres211}, solid (dashed) lines indicate $P\ped{res}>0$ ($P\ped{res}<0$).}
\label{fig:Pres-comparison}
\end{figure*}

The flux of scalar energy to infinity has been identified in \cite{Brito:2023pyl,Dyson:2025dlj} with the nonrelativistic ionization power $P\ped{ion}$, introduced in \cite{Baumann:2021fkf,Baumann:2022pkl,Tomaselli:2023ysb} and defined in \eqref{eqn:Pion}. This is because the ionized, unbound states are the only ones with a nonzero infinity flux. A quantitative comparison between the relativistic and nonrelativistic results was done in \cite{Dyson:2025dlj}, showing satisfactory agreement. Similarly, it seems natural to identify $P\ped{res}$, as defined in \eqref{eqn:Pres}, with the scalar horizon flux of \cite{Brito:2023pyl,Dyson:2025dlj,Li:2025ffh}, because the power lost to excite the bound states is reabsorbed by the BH horizon as they decay. As long as the tidal interaction is treated perturbatively, this reabsorption is much faster than the excitation rate. We now perform a detailed comparison with the relativistic results, kindly shared by the authors of \cite{Brito:2023pyl,Dyson:2025dlj}, which are available up to $R_*=100M$ for $\alpha=0.2$.

We start by comparing the infinity flux and $P\ped{ion}$ in Fig.~\ref{fig:Pion-comparison}. Differently from \cite{Dyson:2025dlj}, we show here a mode-by-mode breakdown in angular components $P\ped{ion}^{\ell m}$, and perform the comparison for retrograde orbits too. The agreement between relativistic and nonrelativistic results is acceptable and improves for large separations, as naturally expected. The curves display a qualitatively similar shape regardless of the mode considered. We note, however, that the comparison is noticeably worse in the retrograde case, with a discrepancy certainly larger than the $\sim\mathcal O(M/R_*)$ relativistic correction naively expected. Modes excited by the dipolar perturbation $\ell_*=1$, i.e.\ $P\ped{ion}^{22}$ and $P\ped{ion}^{00}$, also seem to match worse than the others.

We compare the horizon flux and $P\ped{res}$ in Fig.~\ref{fig:Pres-comparison}. The agreement is in this case significantly worse than it is for ionization. In the prograde case, subdominant modes differ by several orders of magnitude, and even have opposite sign. The dominant $P\ped{res}^{00}$ mode shows instead better agreement, including the peak around $R_*=40M$ due to the $\ket{211}\to\ket{100}$ resonance. The small-$R_*$ behavior of $P\ped{res}^{00}$ is however different, diverging (vanishing) in the nonrelativistic (relativistic) case. The $P\ped{res}^{00}$ mode is entirely mediated by the dipolar perturbation $\ell_*=1$, and the $R_*\to0$ divergence is a direct consequence of the functional form of $V_*$ in \eqref{eqn:V_*}, see also \cite{Tomaselli:2023ysb}. In the retrograde case, subdominant modes also differ by orders of magnitude, but have the same sign. Conversely, the dominant $P\ped{res}^{00}$ mode has opposite sign across most of the domain.

The accumulation of resonances with $\ket{n00}$ states is seen both in the nonrelativistic and relativistic results, at $R_*\gtrsim97M$. We note, however, that their amplitude in the relativistic case is far too small to match the corresponding ionization jump, and remove the associated sharp feature, as would instead be expected from our arguments in Sec.~\ref{sec:no-sharp-features}.

We do not currently know what is the origin of these differences between nonrelativistic and relativistic results. We list here a few speculations.
\begin{enumerate}[label=(\roman*)]
\item The dipolar $\ell_*=1$ component of the perturbation is closely related to the choice of reference frame. In this paper we assumed that the cloud's states $\ket{n\ell m}$ are defined in the accelerated frame of the larger BH, and we included in the potential \eqref{eqn:V_*} the corresponding fictitious force. Defining the unperturbed states in a different frame would impact the components $P\ped{res}^{00}$ and $P\ped{res}^{22}$ (and similarly for $P\ped{ion}$), but not the other ones.
\item Unbound states should also contribute to the horizon flux.\footnote{We thank Rodrigo Vicente for suggesting this idea.} However, we expect their contribution to only be sizeable for the $(\ell,m)=(0,0)$ components. Furthermore, this would be a strictly positive correction, which would then not be able to fix the discrepancy in all cases.
\item When going from \eqref{eqn:Mdot-one-state} to \eqref{eqn:Mdot}, we summed over all Fourier modes, an approximation that is justified as long as one of them is close to resonance. The off-resonance flux, although very subdominant, might then be miscalculated---see the Appendix B of \cite{Takahashi:2024fyq} for a related discussion. Though this caveat should not apply to equatorial orbits, as they only feature one Fourier mode per state.
\item There might be subtle differences between the definition of the relativistic energy fluxes, and that of the nonrelativistic power.
\item There is a very large difference in magnitude between the dominant $P\ped{res}^{00}$ mode and the subdominant $P\ped{res}^{\ell m}$ modes with $\ell\ge1$. It is not clear how reliably the latter are determined in the relativistic case, as the numerical precision required for their calculation is very high.
\item If none of these possible corrections is found to work, one might wonder whether there is something inherently wrong in the nonrelativistic approach, based on bound and unbound states. For example, while these form a complete basis for Hermitian Hamiltonians, the same might not be true for a dissipative system like ours.
\end{enumerate}
We hope to resolve this issue in a future work.

\section{Conclusions}
\label{sec:conclusions}

In light of the scientific potential of next-generation GW observatories, it is crucial to accurately model the interaction of compact binaries with their environment. In this paper, we take a step forward for superradiant clouds, arguably one of the most unique (albeit hypothetical) kinds of BH environment, and certainly among the most interesting ones for the purpose of fundamental physics.

We show that resonances with large decay width can be studied through an effective description that removes the need to solve the Schrödinger equation, even when their orbital backreaction is taken into account. This approach, based on an instantaneous ``resonance power'' $P\ped{res}$, allows for an easier study of certain cases that had not yet been analyzed in the literature. Examples are orbits with generic eccentricity and inclination, as well as the wide floating orbit due to the resonance with the $\ket{100}$ state. In particular, by studying the latter, we show that for orbital inclinations $\iota\lesssim\SI{140}{\degree}$ there is an upper bound to the mass of the cloud that can survive up to the latest stages of the inspiral, which is where the GW signal is loudest.

Perhaps most importantly, computing $P\ped{res}$ also clearly displays the accumulation of infinitely many, increasingly weak resonances at certain orbital frequencies. It had been noticed in the literature that these points corresponded to the discontinuities of the ionization power \cite{Tomaselli:2024bdd,Tomaselli:2024dbw,DellaMonica:2025zby}, but it was not until now that the consequences of this fact were fully appreciated. The sharp features in the binary's evolution and its waveform, predicted from ionization alone, completely disappear when these resonances are properly taken into account.

The claim of sharp features certainly contributed to the belief that superradiant clouds, and similar systems like boson stars \cite{Duque:2023seg}, stood out compared to other BH environments. In light of the results of the present work, it is natural to ask, if not sharp features, then what kind of distinctive features (if any) do boson clouds leave on the orbital waveform. It remains true that, due to the high cloud mass density, the magnitudes of $P\ped{ion}$ and $P\ped{res}$ can overcome $P_\slab{gw}$ in a wide range of parameters. The observation, first pointed out in \cite{Baumann:2022pkl}, that the frequency chirp follows a universal shape that only depends on the state of the cloud (and not, e.g., on $\alpha$), remains therefore true. Such a shape will however not include frequency kinks.

Even with the tools developed here, we are still not quite in a position where we can consistently solve for the orbital evolution in full generality. The reason is that, as discussed in Sec.~\ref{sec:chirping-frequency}, the behavior of resonances changes when their width $\abs{\Gamma_b}$ becomes smaller than the frequency chirp rate $\dot\Omega$. But $\dot\Omega$ in turn depends on the the cloud's properties through $P\ped{res}$ and $P\ped{ion}$, creating a nonlinear feedback. Because most Bohr resonances are non-adiabatic \cite{Tomaselli:2024bdd}, we hope that this subtlety can be resolved in a relatively simple way, for example by iteratively solving for $\Omega(t)$. We also point out that the cancellation of discontinuities shown in Sec.~\ref{sec:no-sharp-features} occurs regardless of whether resonances are wide or narrow.

Our results prompts detectability studies, such as \cite{Cole:2022yzw,DellaMonica:2025zby}, to be re-done by carefully incorporating the effect of accumulating resonances. Furthermore, we anticipate that more work is needed to understand the origin of the discrepancies with the relativistic results. We hope to come back to these questions in a future work.

\section*{Acknowledgements}

G.M.T.~is grateful to Richard Brito, Conor Dyson, and Thomas Spieksma for sharing the data of \cite{Brito:2023pyl,Dyson:2025dlj}, and to Thomas Spieksma, Rodrigo Vicente, and Huiyu Zhu for helpful discussions and comments on the draft. G.M.T.~acknowledges support from the Sivian Fund at the Institute for Advanced Study.

\clearpage
\bibliography{main}% Produces the bibliography via BibTeX.

\end{document}